\documentclass[aps,epsfig,manuscript]{revtex4}
\usepackage{amsmath,amssymb,graphicx}
\usepackage{epsfig}
\usepackage{graphicx}
\usepackage{enumerate}

\newcommand{\avg}[1]{\left< #1 \right>} 
\renewcommand{\d}[2]{\frac{d #1}{d #2}} 
\newcommand{\dd}[2]{\frac{d^2 #1}{d #2^2}} 
\newcommand{\ket}[1]{\left| #1 \right>} 
\newcommand{\intinf}{\int_{-\infty}^\infty}
\newcommand{\beq}{\begin{equation}}
\newcommand{\eeq}{\end{equation}}

\newcommand{\bea}{\begin{eqnarray}}
\newcommand{\eea}{\end{eqnarray}}

\begin{document}



\title{Non-equilibrium spin-boson model: counting statistics and the heat exchange fluctuation theorem}
\author{Lena Nicolin}
\affiliation{Chemical Physics Theory Group, Department of Chemistry, University of Toronto,
80 Saint George St. Toronto, Ontario, Canada M5S 3H6}
\author{Dvira Segal}
\affiliation{Chemical Physics Theory Group, Department of Chemistry,
University of Toronto, 80 Saint George St. Toronto, Ontario, Canada
M5S 3H6}

\date{\today}
\begin{abstract}
We focus on the non-equilibrium two-bath spin-boson model, a toy
model for examining quantum thermal transport in many-body open
systems. Describing the dynamics within the NIBA equations,
applicable, e.g., in the strong system-bath coupling limit and/or at
high temperatures, we derive expressions for the cumulant generating
function in both the markovian and non-markovian limits by
energy-resolving the quantum master equation of the subsystem. For a
markovian bath, we readily demonstrate the validity of a
steady-state heat exchange fluctuation theorem. In the non-markovian
limit a ''weaker" symmetry relation generally holds, a general outcome
of microreversibility.
We discuss the
reduction of this symmetry relation to the universal steady-state
fluctuation theorem. Using the cumulant generating function,
an analytic expression for the heat current is obtained.
Our results establish the validity of the steady-state
heat exchange fluctuation theorem in quantum systems with strong
system-bath interactions. From the practical point of view, this
study provides tools for exploring transport characteristics of the
two-bath spin-boson model, a prototype for a nonlinear thermal
conductor.
\end{abstract}


\maketitle



\section{Introduction}

Impurity models were proved to be extremely useful in predicting
many physical phenomena. The famous {\it spin-boson model}
\cite{Weiss,Legget}, describing the dynamics a single charge on two
states coupled to a dissipative bath, e.g., a solvent, exhibits rich
phenomenology, including various phase transitions. Its applications
range from charge transfer reactions in biological systems
\cite{Marcus}, photosynthesis \cite{Photo}, and the Kondo problem
for magnetic impurities \cite{Kondo}. A variant of the model is the
{\it spin-fermion model}, where a qubit (spin) interacts with one or
more metallic environments \cite{MitraSpin,SMarcus, Sarma}.
These celebrated impurity models are appealing from various reasons.
First, they enclose rich dynamical phenomenology, e.g., the Marcus
theory \cite{Weiss} and the Kondo physics \cite{Kondo}. More
recently, addressing molecular electronic experiments, such generic
models were proved to be useful in predicting various aspects of
molecular transport characteristics \cite{GalperinS}. Secondly, they
serve as a benchmark for developing simulation techniques and
approximation schemes, for describing the dynamics of open many-body
systems \cite{QUAPI,IF}.

The traditional spin-boson (SB) model, considering an impurity-spin
coupled to a {\it single thermal reservoir}, serves as a prototype
model for exploring quantum dissipation problems \cite{Weiss}. The
non-equilibrium version of this model, referring to the case where
the spin (subsystem) is coupled to two thermal reservoirs, has been
suggested as a toy model for exploring quantum transport
phenomenology through an anharmonic nanojunction \cite{Rectif,
SegalM}. In this case, the generic situation is one of a
non-equilibrium steady-state, regardless of the initial preparation.
We refer to this model as the "non-equilibrium spin-boson model"
(NESB).
%
Given the complex dissipative spin dynamics observed in the
single-bath SB model \cite{Weiss}, one expect its non-equilibrium
extension to reveal tangled transport properties. Fundamental topics
of interest are the scaling of the energy current with the spin-bath
coupling strength, the role the reservoirs spectral function and the
tunneling splitting on the subsystem dynamics and the transport
coefficients, and the onset of nonlinear current-temperature bias
characteristics at strong interactions.

The transport behavior of the unbiased (zero magnetic field) NESB
model has been studied perturbatively, under the assumption of weak
system-bath interactions, using master equation methods
\cite{SegalM,Teemu}. While this scheme, providing simple analytic
expressions, can capture some of the aspects of the energy transport process,
the inherent weak system-bath coupling assumption results in a
resonance-sequential transport process where the two reservoirs
separately excite and relax the subsystem. Exact numerical results
can be obtained by following the Keldysh approach \cite{Thoss2} or
by using the complex machinery of the multilayer multiconfiguration
Hartree approach \cite{Thoss1}.

In this paper we present an {\it analytical} study of the NESB model
in the strong coupling limit and/or at high temperatures. 
In this limit a concerted action of the two baths takes place,
where at each relaxation or excitation process both reservoirs contribute in a non-additive manner.
This renders the master equation description complex, 
since the amount of energy transferred between the two baths is no
longer in a one-to-one relationship with the number of spin flip
events.
The objective of our analysis is the cumulant generating function
(CGF). With this at hand, one can derive analytic expressions for the
transport coefficients: the current and its cumulants, exposing
their dependence on the microscopic parameters. Furthermore, given
the CGF, the validity of the steady-state heat exchange fluctuation
theorem \cite{Jarzynski} can be established for anharmonic quantum
models in the strong coupling limit.

The fluctuation theorem (FT) for entropy production quantifies the
probability of negative entropy generation, measuring ''second law
violation" \cite{Evans,Cohen}. Both transient and steady-state
fluctuation theorems (SSFT) have been derived, where the former
looks at non steady-state processes over a finite time $t$, and the
latter measures entropy production in non-equilibrium steady-state
systems over a long interval. In the context of heat exchange
between two equilibrium reservoirs, $\nu=L,R$, the SSFT can be
roughly stated as \cite{Jarzynski,Abe}
\bea \ln [\mathcal P_{t}(+\omega)/\mathcal P_t(-\omega)]=\Delta
\beta \omega. \label{eq:SSFT} \eea
Here $\mathcal P_t(\omega)$ denotes the probability
distribution of the net heat transfer $\omega$, from $L$ to $R$,
over the (long) interval $t$, with $\Delta \beta=T_R^{-1}-T_L^{-1}$
as the difference between the inverse temperatures of the reservoirs.
Extending the work and heat FT to the quantum domain has recently
attracted significant attention \cite{Mukamel-rev, Hanggi-rev}.
Specifically, a quantum exchange FT, for the transfer of energy
between two reservoirs maintained at different temperatures, has
been derived in Refs. \cite{Jarzynski, Tasaki, Campisi} using
projective measurements, and in Refs. \cite{Mukamel-weak,
HanggiBerry}, based on the unraveling of the quantum master equation
(QME). These derivations assume that the interaction between the two
thermal baths is weak, and can be neglected with respect to overall
energy changes. Using the Keldysh approach, an exact analysis was
carried out in Ref. \cite{Dhar}. However, it is valid only for
harmonic systems. It is thus an open question whether a heat
exchange FT is obeyed by an anharmonic quantum system {\it strongly
coupled} to multiple reservoirs.

Another subtle point is the role of non-markovian effects on the
heat exchange SSFT and the current cumulants. In charge transfer
problems, this topic has recently attracted significant interest
\cite{Antti,Emary1,Emary2}. The analogous problem, the reflection of
non-markovian effects within the CGF in energy exchange scenarios
has been considered for equilibrium systems in Ref. \cite{Hatano}.
The Markov approximation is justified once the relaxation of the
bath is fast, while the dynamics of the subsystem is slow. In this case
the amount of energy transferred between the subsystem and the bath is
pinned down with an arbitrary precision, as a strict energy
conservation condition is enforced. However, once the assumption of
markovianity is relaxed, when the dynamics of the baths degrees of freedom is on a
comparable timescale with the subsystem evolution,
energy-non-conserving processes on short time scales due to the
energy-time uncertainty (when looking only at a subsystem) cannot be
excluded. On this bath-decorrelation time scale, it is  not obvious
that the basic symmetry [Eq. (\ref{eq:SSFT})] still holds. 

Considering the NESB model in the strong interaction limit, allowing
for non-markovian effects, it is our objective here to investigate
its heat exchange properties: (i) To obtain the CGF and gain
an explicit expression for the heat current,
useful for understanding heat current characteristics for
anharmonic-strongly coupled systems. (ii) Given the CGF, to derive
the heat exchange SSFT. (iii) To understand the role of
non-markovian (memory) effects on the onset of the SSFT.
%
Our analysis makes use of the noninteracting-blip approximation
(NIBA) \cite{Weiss}. This scheme can faithfully simulate the SB
dynamics at strong system-bath interactions and/or at high
temperatures in the Ohmic case. It is also exact for the unbiased case
at weak damping.
Under this approximation, the subsystem's dynamics is
described within a time convolution quantum master equation. We
unravel this dynamical equations into trajectories with a particular
amount of net energy dissipated at each contact. In the markovian
limit a heat exchange SSFT is verified. We also obtain the CGF,
independent of the particular physical realization. In the
non-markovian case a symmetry relation is recovered
\cite{Hanggi-rev}, reaching the universal SSFT once the observation
time $t$ [Eq. (\ref{eq:SSFT})] is much greater than the bath
decorrelation time.

The paper is organized as follows. In Sec. II, we describe our model
and recall known results for the spin-boson model in the strong
coupling limit. Sec. III presents results for the CGF in the
markovian limit, introducing the concepts and definitions that will
become useful once the more involved non-markovian case is
considered in Sec. IV. In Sec. V we conclude.


\section{Model and dynamics}

The non-equilibrium spin-boson Hamiltonian, comprising a spin
subsystem coupled to two ($\nu=L,R$) independent phonon baths,
maintained at a temperature $T_{\nu}$, is described by the
Hamiltonian ($\hbar\equiv 1$)
\beq
H = \frac{\omega_0}{2} \sigma_z +  \frac{\Delta}{2}\sigma_x +
\sigma_z \sum_{\nu,j}\lambda_{j,\nu}(b_{j,\nu}^\dagger + b_{j,\nu}) +
\sum_{\nu,j}\omega_j b_{j,\nu}^{\dagger}b_{j,\nu}.
\label{eq:HSB}
\eeq
Here $\sigma_x$ and $\sigma_z$ are the Pauli matrices, $\omega_0$ is
the energy gap between the spin levels, and $\Delta$ is the
tunneling energy. Explicitly, in the two-state basis,
$\sigma_z=|1\rangle\langle 1| -|0\rangle\langle 0|$ and
$\sigma_x=|1\rangle\langle 0| +|0\rangle\langle 1|$. Each reservoir
includes a collection of uncoupled harmonic oscillators,
$b_{j,\nu}^{\dagger}$ ($b_{j,\nu}$) is the bosonic creation
(annihilation) operator of the mode $j$ in the $\nu$ reservoir. The
parameter $\lambda_{j,\nu}$ accounts for the system-bath interaction
strength.



The transport characteristic of the non-equilibrium spin-boson model
can be obtained exactly using numerical simulations \cite{Thoss1}.
Here, with the motivation to gain insight into the heat current
characteristics, the behavior of the current cumulants, and the
fluctuation symmetries we resort to approximations, allowing for
analytical results. In particular, we employ the NIBA equations,
valid at strong system-bath interactions or for high temperatures,
assuming an Ohmic spectral density \cite{Legget, Weiss}. The NIBA
equations can be also obtained by applying the Born approximation
with respect to the dressed tunneling elements \cite{Dekker,
Aslangul}. While this method has been originally derived for a spin
coupled to a {\it single} bosonic reservoir, one can trivially
generalize it to describe a multi-bath case.

We begin by transforming the SB Hamiltonian (\ref{eq:HSB}) to the
displaced bath-oscillators basis using the small polaron
transformation \cite{Mahan}, $H_p=U^{\dagger}HU$,
$U=e^{i\sigma_z\Omega/2}$,
%
\bea
H_{p} = \frac{\omega_0}{2} \sigma_z +
\frac{\Delta}{2} \left( \sigma_+ e^{i\Omega} + \sigma_- e^{-i\Omega} \right)
+\sum_{\nu,j}\omega_j b_{j,\nu}^{\dagger}b_{j,\nu},
\label{eq:HSBs}
\eea
where $\sigma_{\pm}=\frac{1}{2}(\sigma_x\pm i \sigma_y)$, or
$\sigma_+=|0\rangle \langle 1|$, $\sigma_-=|1\rangle \langle 0|$,
are the auxiliary Pauli matrices, $\Omega=\sum_{\nu}\Omega_{\nu}$,
and
$\Omega_{\nu}=2i\sum_{j}\frac{\lambda_{j,\nu}}{\omega_{j}}(b_{j,\nu}^{\dagger}-b_{j,\nu})$.
Under the NIBA approximation \cite{Legget,Dekker,Aslangul},
generalized to the two-baths case, it can be shown that the spin
polarization $\langle \sigma_z(t) \rangle $ obeys a convolution-type
master equation
\bea
\frac{d \langle \sigma_z\rangle}{dt}= -\int_{0}^{t}
K_{s}(t-\tau)\langle \sigma_z(\tau)\rangle d\tau -\int_{0}^{t}
K_a(t-\tau)d\tau,
\label{eq:integro}
\eea
where the symmetric and antisymmetric kernels are given by
\bea K_s(t)=\Delta^2e^{-Q'(t)}\cos[Q''(t)]\cos(\omega_0 t)
\nonumber\\
K_a(t)=\Delta^2e^{-Q'(t)}\sin[Q''(t)]\sin(\omega_0 t). \label{eq:K}
\eea
The complex function $Q(t)=\sum_{\nu}{Q_{\nu}(t)}$, made of a real
and imaginary components, $Q_{\nu}(t)=Q'_{\nu}(t)+iQ''_{\nu}(t)$, is defined
by the correlation function $e^{-Q(t)}=\langle e^{i\Omega(t)}e^{-i\Omega(0)}\rangle$,
with the thermal average performed over both reservoirs degrees of freedom.
It is given by
\bea
Q''_{\nu}(t)& = &  \int_{0}^{\infty} \frac{J_{\nu}(\omega)}{\pi\omega^2}\sin(\omega t)d\omega,
\nonumber\\
Q'_{\nu}(t)& = & \int_{0}^{\infty}\frac{J_{\nu}(\omega)}{\pi\omega^2}[1-\cos(\omega t)] [1+2n_{\nu}(\omega)] d\omega.
\label{eq:Q}
\eea
Here $J_{\nu}(\omega)$ is the $\nu$-bath spectral function,
incorporating system-bath interactions
\bea
J_{\nu}(\omega)=4\pi\sum_{j}\lambda_{j,\nu}^2\delta(\omega-\omega_j).
\label{eq:spectral}
\eea
In what follows we focus on the two-state population dynamics,
therefore we rewrite Eq. (\ref{eq:integro}) in terms of the states
population
\bea \frac{dp_1(t)}{dt}&=& -\frac{\Delta^2}{2} \int_{0}^{t}
e^{-Q'(t-s)} \cos[ \omega_0(t-s)-Q''(t-s)] p_1(s) ds
\nonumber\\
& +&\frac{\Delta^2}{2} \int_{0}^{t} e^{-Q'(t-s)} \cos[
\omega_0(t-s)+Q''(t-s)] p_0(s)ds,
\nonumber\\
1&=&p_0(t)+p_1(t),
\label{eq:integrP}
 \eea
where $\avg{\sigma_z(t)}=p_1(t)-p_0(t)$. We explore next the heat
transport characteristics under the NIBA approximation (i) assuming
a markovian dynamics,
and (ii) more generally, retracting to the non-markovian case,
allowing for memory effects in the thermal baths. The non-markovian
analysis can be reduced to the markovian description in the
appropriate limit. For clarity, we have decided to first present
here the (simple) markovian limit, then generalize the analysis and
portray the non-markovian regime. This allows us to introduce the
main concepts involved in the CGF derivation within a relatively
simple setup.

\section{Markovian limit}

\subsection{Population Dynamics}

A general analysis of counting statistics of a
 multi-state system connecting two non-equilibrium markovian reservoirs
has been carried out in Ref. \cite{FTshort}, based upon the NIBA
equations. We use this scheme and derive here the CGF for the NESB
model. In the markovian limit one assumes that the spin system
slowly evolves in comparison to the reservoirs evolution. Thus, we
make the following two simplifications in the integro-differential
equation (\ref{eq:integrP}): First, we replace the population,
$p_n(s)$ by $p_n(t)$ ($n=0,1$), supposing that the timescale over
which the memory, represented by the integral, is important, is
sufficiently short. Second, we extend the integral upper limit to
infinity, assuming the integrand quickly dies out. Under these
approximations, Eq. (\ref{eq:integrP}) reduces to a kinetic equation
for the population dynamics,
\bea \dot p_1=-k_dp_1(t) +k_up_0(t).
\label{eq:EOMsm} \eea
The rate constants are given as Fourier transforms of bath
correlation functions,
\bea
k_d =
C(\omega_0),\,\,\,\,\,\,
k_u =
C(-\omega_0),
\eea
with
\bea
C(\omega_0) =  
\int_{-\infty}^\infty
e^{i\omega_0 t} C_L(t) C_R(t) dt.
\label{eq:Cw0}
\eea
The ingredients of this correlation function are given in terms of the function
$Q_{\nu}(t)$, defined in Eq. (\ref{eq:Q}),
\bea
C_{\nu}(t)=
\frac{\Delta}{2}
e^{-Q_{\nu}(t)}.
\label{eq:Ct}
\eea
Using the convolution theorem, the transition rates $C(\pm
\omega_0)$ can be rewritten as a convolution of the $L$-bath and
$R$-bath induced processes,
\bea C(\omega_0)&  = & \frac{1}{2\pi}\int_{-\infty}^\infty
C_L(\omega_0 - \omega)C_R(\omega) d\omega, \label{eq:conv} \eea
introducing the Fourier transform
\bea
C_{\nu}(\omega)=\int_{-\infty}^{\infty}e^{i\omega t}C_{\nu}(t)dt.
\label{eq:Cnu}
\eea
These bath-specific microscopic rates satisfy the detailed balance
relation,
\bea
\frac{C_{\nu}(\omega)}{C_{\nu}(-\omega)}&  = & e^{\omega\beta_{\nu}}.
\label{eq:DB}
\eea
However, such a relation does not hold for the combined rate
$C(\omega)$, ruling the dynamics. The QME (\ref{eq:EOMsm}) encloses
complex physical processes as Eq. (\ref{eq:conv}) draws nontrivial
transfer rates. When the system decays it disposes the energy
$\omega_{0}$ into both reservoirs, cooperatively; the energy $\omega$
is dissipated into the $R$ bath while the $L$ bath gains (or
contributes) the rest, $\omega_{0}-\omega$. Similarly, excitation of
the system occurs through an $L$-$R$ compound process. Since energy
is dissipated or absorbed in such complex processes, energy
"counting" is a nontrivial task, as reflected in the resolved master
equation (\ref{eq:Pw}) discussed below.

\subsection{Cumulant Generating Function}

We construct next the cumulant generating function for the NESB
model in the NIBA-markovian limit presented above. Following Ref.
\cite{HanggiBerry}, we begin by defining the function
$\mathcal P_t(n,\omega)$ as the probability that within the time $t$ a total
of energy $\omega$ has been transferred from the left bath to the
right bath, while the spin is populating the $n$ ($n=0,1$) state at
time $t$. The time evolution of this quantity follows
\bea \frac{d \mathcal P_t(0,\omega)}{dt}& = & - \mathcal
P_t(0,\omega) \intinf\frac{1}{2\pi}
C_R(-\tilde{\omega})C_L(\tilde{\omega}-\omega_0) \,d\tilde{\omega}
\nonumber\\
&+& \intinf \frac{1}{2\pi} C_R(\omega-\tilde{\omega})C_L(\omega_0- (\omega - \tilde{\omega}))\mathcal P_t(1,\tilde{\omega})  \,d\tilde{\omega} \nonumber \\
\frac{d \mathcal P_t(1,\omega)}{d t} & = & - \mathcal P_t(1,\omega)
\intinf \frac{1}{2\pi} C_R(\omega)C_L(\omega_0-\omega) \,d\omega
\nonumber\\
&+& \intinf \frac{1}{2\pi}C_R(\omega-\tilde{\omega})C_L(
(\tilde{\omega} -\omega)-\omega_0)\mathcal P_t(0,\tilde{\omega})
\,d\tilde{\omega}, \label{eq:Pw} \eea
for details see Appendix A. One can rationalize this equation as
follows. Focusing for example on the dynamics of $\mathcal
P_t(1,\omega)$, the first term in this rate equation describes the
decay of this probability as the spin flips to the ground state and
extra energy is dissipated to the $R$ reservoir. The second term
collects processes with an energy $\tilde \omega$ transferred to the
$R$ bath by the time $t$, with the spin occupying the ground state. At time
$t$ a spin flip takes place accompanied by an extra energy
$\omega-\tilde \omega$ dissipated to the $R$ bath, completing the transfer of a total amount
of energy $\omega$ to the right bath at $t$.

We note that in the present model we cannot adopt standard approaches for unraveling
the reduced density matrix,  directly dressing the
interaction term in the Hamiltonian by the counting process \cite{Mukamel-rev}. This in because the two reservoirs affect
the energy transfer process in a nonlinear way, thus counting system-bath interaction processes
(as in a perturbation theory series) does not reveal here the actual amount of energy exchanged between
the two reservoirs.

We Fourier transform the above system of equations to obtain the
{\it characteristic function} $Z(\chi,t)$ for the energy counting
field $\chi$,
\bea
\ket{Z(\chi,t)} \equiv
\begin{pmatrix}
\intinf \mathcal P_t(0,\omega)e^{i\omega\chi}\,d\omega \\\intinf
\mathcal P_t(1,\omega)e^{i\omega\chi}\,d\omega
\end{pmatrix}
\label{eq:z}
\eea
It satisfies the differential equation 
\beq \d{\ket{Z(\chi, t)}}{t} = - \hat W(\chi)\ket{Z(\chi, t)},
\label{eq:Z} \eeq
where the matrix $\hat W$ contains the following elements
\bea \hat W(\chi) =
\begin{pmatrix}
C(-\omega_0) & -C^d(\chi)  \\
-C^u(\chi) & C(\omega_0)  \\
\end{pmatrix}
\label{eq:mu}
\eea
The diagonal terms were defined above, see Eq. (\ref{eq:Cw0}). The
nondiagonal terms are given by the integrals
\bea C^{d/u}(\chi) = \frac{1}{2\pi}\intinf
C^{d/u}(\omega)e^{i\omega\chi}  d\omega \label{eq:Cdu} \eea
with the components
\bea
&& C^{d}(\omega) = C_R(\omega)C_L(\omega_0- \omega )
\nonumber\\
&& C^{u}(\omega) =C_R(\omega)C_L( - \omega-\omega_0).
\label{eq:Cdu2}
\eea
The {\it cumulant generating function} is formally defined as
\bea G(\chi)= \lim_{t \to \infty} \ \frac{1}{t}\ln \intinf \mathcal
P_t(\omega)e^{i\omega\chi}d\omega, \label{eq:Gd2} \eea
where we introduced the short notation, ${\mathcal
P}_{t}(\omega)=P_t(0,\omega)+P_t(1,\omega)$, the probability to
transfer by the time $t$ an energy $\omega$ from left to right,
irrespective of the spin state. In the present case the CGF is
expressed in terms of $|Z\rangle$ as
\bea
G(\chi) =  \lim_{t \to \infty} \ \frac{1}{t}\ln \langle I| Z(\chi,t) \rangle,
\eea
with $\langle I|= \langle 1 1|$, denoting a left vector of unity. It
is practically given by the negative of the smallest eigenvalue of
the matrix $\hat W$. We diagonalize $\hat W$ and explicitly obtain
the CGF in terms of the microscopic rates,
\bea
G(\chi)= -\frac{C(\omega_0)+C(-\omega_0)}{2} + \frac{\left[  (C(\omega_0)-C(-\omega_0))^2+4C^d(\chi)C^u(\chi) \right]^{1/2}}{2}.
\label{eq:G}
\eea
The heat current and its noise power can be readily derived, by taking the first and the second derivatives,
respectively, of the CGF
\bea
&&\langle J\rangle \equiv \frac{\avg{\omega}_{t}}{t} = \d{G(\chi)}{(i\chi)}\Big|_{\chi=0}, \,\,\,\,\,
\nonumber\\
&&\avg{S}  \equiv \frac{\avg{\omega^2}_{t} - \avg{\omega}^2_{t}}{t} =  \dd {G(\chi)}{(i\chi)}\Big|_{\chi=0}.
\label{eq:J}
\eea
Here $\avg{\omega}_{t}$ denotes the total energy $\omega$ transferred from $L$ to $R$ by the (infinitely long) time $t$.
Using the formal structure (\ref{eq:G}) one can show that the steady-state
heat current, defined as positive when flowing left to right, obeys
\bea \avg{J}=\frac{1}{2\pi}\int_{-\infty}^{\infty}\omega d\omega
\left[ C_R(\omega)C_L(\omega_0-\omega)p_1 -
C_R(-\omega)C_L(-\omega_0+\omega)p_0 \right]. \label{eq:J2} \eea
This expression incorporates the steady-state populations \bea
p_1=C(-\omega_0)/(C(\omega_0)+C(-\omega_0)), \,\,\,\,
p_0=C(\omega_0)/(C(\omega_0)+C(-\omega_0)). \label{eq:SSpop} \eea
For details see Appendix B. The result for the heat current agrees
with the expression used ad-hoc in Refs. \cite{Rectif, SegalM}. It
can be rationalized by viewing $\omega
C_R(\omega)C_L(\omega_0-\omega)p_1$ as a spin relaxation process
with  the energy $\omega$ directed to the $R$ bath and the amount of
$\omega_0-\omega$ disposed into the $L$ bath. Similarly, the second
term describes energy loss from the $R$ bath, where, combined with
an energy influx from the $L$ bath, results in the excitation of the
spin system. It is significant to note that this expression has been
achieved under relatively general conditions, for systems satisfying
a markovian-NIBA approximation. The details of the Kernel
$K_{s/a}(t)$ (e.g., the bath statistics) are not utilized in this
derivation. Thus, it is valid for other systems following the
structure (\ref{eq:EOMsm})-(\ref{eq:Cw0}), e.g., the spin-fermion
model \cite{MitraSpin,SMarcus,FTshort}. Appendix B further details
the derivation of the the second cumulant, the noise power of the
NESB junction,
\bea \langle S \rangle  &= &
p_1\int_{-\infty}^{\infty}\frac{1}{2\pi}\omega^2C_R(\omega)C_L(\omega_0-\omega)d\omega
+
p_0\int_{-\infty}^{\infty}\frac{1}{2\pi}\omega^2C_R(-\omega)C_L(\omega-\omega_0)d\omega+
\nonumber\\
&-&
2\frac{1}{C(\omega_0)+C(-\omega_0)}\frac{1}{(2\pi)^2}\int_{-\infty}^{\infty}\omega
C_R(-\omega)C_L(\omega-\omega_0)d\omega\int_{-\infty}^{\infty}\omega
C_R(\omega)C_L(\omega_0-\omega)d\omega+
\nonumber\\
&-& 2\frac{1}{C(\omega_0)+C(-\omega_0)}\avg{J}^2.
\label{eq:S2}
\eea
%

\subsection{Fluctuation Theorem}

We continue and confirm the validity of the SSFT in the NESB model,
under the Markov approximation. This relation can be established by
examining the symmetry of the CGF, Eq. (\ref{eq:G})
\cite{Mukamel-rev}. It is clear that it is sufficient to focus on
the product term, $\mathcal D(\chi)\equiv C^d(\chi) C^u(\chi)$, for
resolving the symmetry of $G(\chi)$. Using the definitions
(\ref{eq:Cdu})-(\ref{eq:Cdu2}) we therefore write (ignoring the
$2\pi$ prefactors)
\bea
\mathcal D(\chi)=
\int_{-\infty}^{\infty}e^{i\omega\chi} C_R(\omega)C_L(\omega_0-\omega)d\omega \times
\int_{-\infty}^{\infty}e^{i\omega\chi} C_R(\omega)C_L(-\omega_0-\omega)d\omega.
\eea
Shifting the argument $\chi\rightarrow (i\Delta \beta-\chi)$,
$\Delta\beta=\beta_R-\beta_L$, it translates to
\bea
\mathcal D(i\Delta \beta-\chi)
&=& \int_{-\infty}^{\infty}e^{-i\omega\chi}e^{-\omega \Delta \beta} C_R(\omega)C_L(\omega_0-\omega)d\omega
\nonumber\\
&\times&
 \int_{-\infty}^{\infty}e^{-i\omega\chi}e^{-\omega\Delta \beta} C_R(\omega)C_L(-\omega_0-\omega)d\omega.
 \label{eq:Prd}
\eea
We now change variables, $\omega\rightarrow -\omega$, then use the
detailed balance relation for $C_{\nu}(\omega)$, see Eq.
(\ref{eq:DB}). This transforms the first element in the RHS of the
above equation to
\bea
&&C^d(i \Delta \beta-\chi)=
\int_{-\infty}^{\infty}e^{i\omega\chi}e^{\omega \Delta \beta} C_R(-\omega)C_L(\omega_0+\omega)d\omega
\nonumber\\
&&=
\int_{-\infty}^{\infty}e^{i\omega\chi}e^{\omega \Delta \beta} C_R(\omega)e^{-\beta_R\omega}C_L(-\omega_0-\omega)
e^{\beta_L(\omega_0+\omega)}d\omega
\nonumber\\
&&=
e^{\beta_L\omega_0} C^u(\chi).
\eea
Similarly, the second element in the RHS of Eq. (\ref{eq:Prd}) reduces to
\bea
&& C^u(i\Delta\beta-\chi)=
\int_{-\infty}^{\infty}e^{i\omega\chi}e^{\omega \Delta \beta} C_R(-\omega)C_L(\omega-\omega_0)d\omega
\nonumber\\
&&=
\int_{-\infty}^{\infty}e^{i\omega\chi}e^{\omega \Delta \beta} C_R(\omega)e^{-\beta_R\omega}C_L(\omega_0-\omega)
e^{-\beta_L(\omega_0-\omega)}d\omega
\nonumber\\
&&=
e^{-\beta_L\omega_0} C^d(\chi).
\eea
Joining these two pieces we conclude that
\bea
C^d(\chi)
C^u(\chi)=  C^d(i \Delta \beta-\chi) C^u(i\Delta\beta-\chi).
\label{eq:symmGF}
\eea
Therefore, the CGF overall satisfies
\bea
G(\chi)=G(i\Delta \beta-\chi).
\label{eq:FT}
\eea
We are now in position to demonstrate the validity of a fluctuation
relation for this non-equilibrium strongly coupled system. The
probability to transfer the energy $\omega$ by the time $t$, from
$L$ to $R$ is given by the inverse Fourier transform of Eq.
(\ref{eq:Gd2}),
\bea {\mathcal P}_{t}(\omega)=\frac{1}{2\pi}\int_{-\infty}^{\infty}
e^{t G(\chi)}e^{-i\omega\chi} d\chi. \label{eq:P} \eea
Similarly, the quantity ${\mathcal P}_{t}(-\omega)$ represents the
probability that overall an energy $\omega$ has been transmitted in
the opposite direction, right to left, up to time $t$.
%
%
Based on the symmetry of the CGF, Eq. (\ref{eq:symmGF}), one readily concludes that 
\bea \lim_{t \to \infty} \ \frac{1}{t} \ln \frac{{\mathcal
P}_{t}(\omega)}{{\mathcal P}_{t}(-\omega)}=\omega \Delta\beta.
\label{eq:SSFTM} \eea
This expression describes a fluctuation relation for the
non-equilibrium SB model, valid beyond the weak-coupling
approximation \cite{Mukamel-weak, HanggiBerry}. Comparing this
result to the weak coupling limit, described in Appendix C, we
observe that formally these two expressions are identical. However,
one should note that in the strong coupling limit the energy
variable $\omega$ is {\it continuous}, since multi-phonon processes
in which part of the energy goes to $L$ bath and part goes to the
$R$ baths, are allowed. In contrast, in the weak coupling limit
energy transfer processes take place in {\it integer} units of the
spin spacing, since this energy travels to either reservoirs
separately.

\subsection{Examples and the Gaussian-Marcus limit}

\begin{figure}[htbp]
\vspace{0mm} \hspace{0mm} {\hbox{\epsfxsize=75mm \epsffile{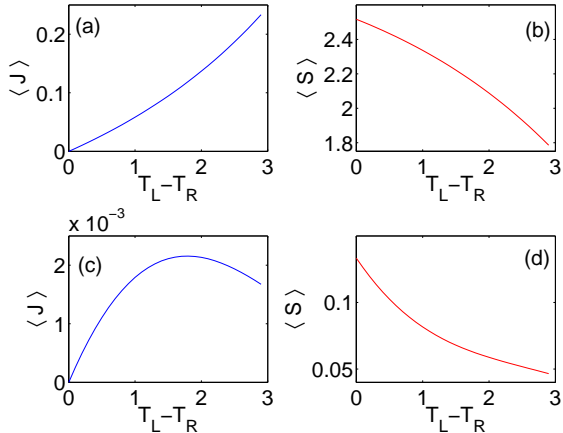}}}
\caption{Energy current and noise power of the spin-boson model in
the Marcus limit, Eqs. (\ref{eq:gauss})-(\ref{eq:gaussC}). Other
parameters are $T_L=5$, $T_R=T_L-\Delta T$, $\omega_0=0.5$,
$\Delta/2=1$, (a)-(b) $E_r^{\nu}=1$, (c)-(d) $E_r^{\nu}=50$.}
\label{FigJS}
\end{figure}

We exemplify our results, and work out an expression for the heat current and the noise power
for a specific case, the so called ``Marcus" limit \cite{Marcus},
assuming high temperatures $T_{\nu}>\omega_0$ and strong coupling.
This limit is reached by performing a short time expansion of $Q(t)$, [Eq. (\ref{eq:Q})] resulting in
\bea Q'_{\nu}(t)=E_r^{\nu}T_{\nu}t^2, \,\,\,\,\
Q_{\nu}''(t)=E_r^{\nu} t. \eea
The reorganization energy
$E_r^{\nu}=\sum_{j}4\lambda_{j,\nu}^2/\omega_j$ incorporates
system-bath interactions. It can be
equivalently expressed in terms of the spectral density
(\ref{eq:spectral}), $E_r^{\nu}=\int \frac{J_{\nu}(\omega)}{\pi
\omega}d\omega$. Using these expressions, the Fourier transform of
the time dependent rates (\ref{eq:Cw0}) and (\ref{eq:Ct}) can be
resolved,
\bea &&C_{\nu}(\omega) =
\frac{\Delta}{2}\sqrt{\frac{\pi}{E_r^{\nu}T_{\nu}}}\exp\left[-\frac{(\omega
- E_r^{\nu})^2}{4T_{\nu}E_r^{\nu}}\right],
\nonumber\\
&&C(\omega_0)= \frac{\Delta^2}{4} \sqrt { \frac{\pi}{E_r^{R}T_R
+E_r^LT_L }} \exp\left[-\frac{(\omega_0 - E_r^L - E_r^R)^2} {4
(T_RE_r^R +T_L E_r^L)}\right]. \label{eq:gauss} \eea
Following Eq. (\ref{eq:J2}), the average heat current can be
analytically obtained
\beq \avg{J} = \Delta^2\frac{\sqrt{2\pi}E_r^LE_r^R\Delta T}{(2E_r^L
T_L+2E_r^R T_R)^{\frac{3}{2}}} \exp\left[-\frac{(E_r^L + E_r^R -
\omega_0)^{2} }{4(E_r^L T_L+E_r^R T_R)}\right]\times f_A
\label{eq:gaussJ}
\eeq
where $f_A = \left\{ \exp\left[\frac{\omega_0(E_r^L + E_r^R)}{(E_r^L
T_L+E_r^R T_R)}\right] + 1\right\}^{-1}$. This result agrees with
\cite{Rectif,SegalM}. The second moment of the current can be
similarly calculated, however the expression is too cumbersome to be
included. We present the behavior of the current and its noise power
in Fig. \ref{FigJS}. We observe nonlinear effects in the energy
current, including the effect of negative differential conductance
\cite{Rectif,SegalM}. The noise drops with increasing bias
temperature.

\begin{figure}[htbp]
\vspace{0mm} \hspace{0mm} {\hbox{\epsfxsize=75mm \epsffile{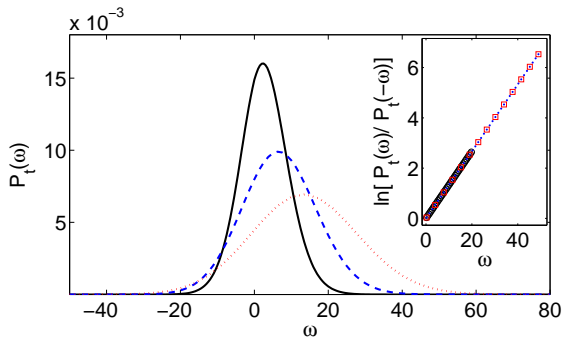}}}
\caption{The probability distribution $\mathcal{P}_{\tau}(\omega)$
at various times, $t=20$ (full), $t=50$ (dashed), $t=100$ (dotted).
Other parameters are
$T_L=5$, $T_R=3$, $E_r^{\nu}=1$, $\omega_0=0.5$, $\Delta/2=1$.
Data was generated using the Marcus rates.
The inset demonstrates for the same data the validity of the fluctuation
theorem, with the slope of $\Delta \beta=0.133$, $t=20$ ($\circ$),
$t=50$ (dotted), $t=100$ ($\square$). } \label{FigP}
\end{figure}

One can in principle seek to derive an analytic form for the
probability distribution function $\mathcal P_t(\omega)$, since the
analytic structure of $\hat W$ is known: The diagonal terms are the
rates, (\ref{eq:gauss}), the nondiagonal part is given by
\bea C^{d/u}(\chi) &=& \frac{\Delta^2}{4}\sqrt{\frac{\pi}{E_LT_L +
E_RT_R}}e^{-\frac{E_L^2 + E_R^2+\omega_0^2 + 2E_L\left[E_R(1-2i\chi
\Delta T+2\chi^2T_LT_R)\mp\omega_0\right] -2E_R\omega_0 \mp
4iE_R\chi T_R\omega_0 }{4(E_LT_L + E_RT_R)}}.
\label{eq:gaussC}
\eea
Since the result is very complex, we retreat here to numerical
simulations. We plug these expressions into the formal solution
(\ref{eq:G}) and perform a numerical inverse Fourier transform [Eq.
(\ref{eq:P})], to obtain the distribution $\mathcal P_t(\omega)$.
Fig. \ref{FigP} demonstrates its shape at different times. The
averaged current and noise agree with the values provided in Fig.
\ref{FigJS}. We have also confirmed that the different curves indeed
satisfy the SSFT (inset).

\begin{figure}[htbp]
\vspace{0mm} \hspace{0mm} {\hbox{\epsfxsize=85mm \epsffile{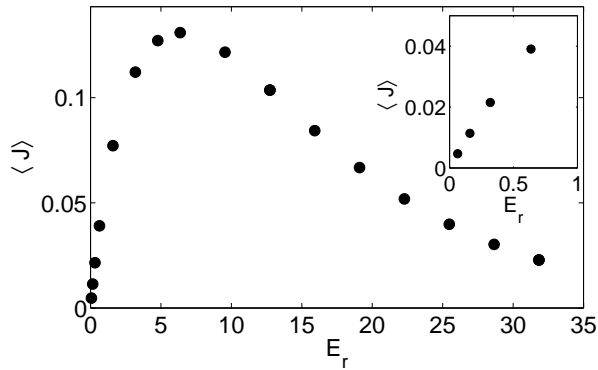}}}
\caption{Energy current in the unbiased spin-boson model, $T_L=5$,
$T_R=3$, $\omega_0=0$, $\omega_c=10$, $\Delta/2=1$, numerically
simulating (\ref{eq:Q}) and the resulting current (\ref{eq:J2}). The
inset zooms on the weak coupling limit, displaying a linear
dependency of the heat current on the reorganization energy. }
\label{FigE}
\end{figure}

We now go beyond the Marcus limit, and demonstrate the behavior of
the current with the reorganization energy $E_r$, quantifying
system-bath coupling strength. We consider the unbiased case
$\omega_0=0$, and take an Ohmic spectral function
$J_{\nu}(\omega)=\frac{\pi E_r^{\nu}}{\omega_c}\omega
e^{-\omega/\omega_c}$, identical for the two baths. Since we are
interested in the energy current behavior for both weak and strong
coupling strengths, we numerically calculate the elements in Eq.
(\ref{eq:J2}) using the definitions (\ref{eq:Q}) and (\ref{eq:Cnu}).
The results are displayed in Fig. \ref{FigE} showing a turnover
behavior, where the current decays with increasing $E_r$, at large
values. It can be easily proved that the weak coupling scheme can
only produce a linear dependence of the current on the coupling
strength \cite{Rectif,SegalM}, see also Eq. (\ref{eq:ACJ}). The
decaying behavior observed here is thus a fingerprint of the strong
coupling limit. Similar results were reported in \cite{Thoss1},
using exact numerical simulations. Practically, this turnover
behavior indicates that for maximizing the rate of energy transport
in nanodevices one should work at the intermediate system-bath
coupling limit.


\section{Non-Markovian Dynamics}

\subsection{Cumulant Generating Function}

We generalize here the results of the markovian analysis and derive
the CGF for the non-markovian model introduced in Sec. II. A
systematic formalism for analyzing non-markovian effects in {\it
charge transfer} systems has been detailed in Refs.
\cite{Antti,Emary1,Emary2}. Here we adapt this scheme to describe
energy transfer processes. Further, while a single counting field
has been introduced in \cite{Antti}, in the present model one should
introduce two such fields, independently counting energy
transmission at each contact.
We begin the analysis by rewriting the equation of motion for the
spin population (\ref{eq:integrP}) as
\bea \frac{dp_1(t)}{dt}&=& -\frac{\Delta^2}{2} \Re \int_{0}^{t}
e^{i \omega_0(t-s)} e^{-Q_L(t-s)}   e^{-Q_R(t-s)}
p_1(s) ds
\nonumber\\
& +&\frac{\Delta^2}{2} \Re \int_{0}^{t}
e^{i\omega_0(t-s)} e^{-Q_L(s-t)} e^{-Q_R(s-t)}  p_0(s) ds,
\nonumber\\
1&=&p_0(t)+p_1(t),
\label{eq:integrPNM}
 \eea
where we made use of the symmetry properties of the $Q(t)$ function,
$Q'(t-s)=Q'(s-t)$ and $Q''(t-s)=-Q''(s-t)$, see the explicit expressions in Eq.
(\ref{eq:Q}). Here $\Re$ denotes the real part. In the next step, we
use the Fourier transform relation
\bea \frac{\Delta}{2}e^{-Q_{\nu}(t)}=\frac{1}{2\pi} \intinf
e^{-i\omega t}C_{\nu}(\omega)d\omega, \eea
and write
\bea \frac{dp_1(t)}{dt}&=& -\frac{1}{2\pi^2} \Re \int_{0}^{t} ds
p_1(s) e^{i \omega_0(t-s)} \intinf d\omega_1 C_L(\omega_1)
e^{-i\omega_1(t-s)} \intinf d\omega_2C_R(\omega_2)
e^{-i\omega_2(t-s)}
\nonumber\\
& +&\frac{1}{2\pi^2} \Re \int_{0}^{t} ds p_0(s) e^{-i\omega_0(s-t)}
\intinf d\omega_1 C_L(\omega_1) e^{-i\omega_1(s-t)} \intinf
d\omega_2 C_R(\omega_2) e^{-i\omega_2(s-t)} . \nonumber\\ \eea
%
We now energy-resolve this equation, $p_n(t)=\intinf
d\omega_L\intinf d\omega_R \mathcal P_t(n,\omega_L,\omega_R)$,
looking for the probability $\mathcal P_t(n,\omega_L,\omega_R)$
at the time $t$ the spin occupies the $n$ ($n=0,1$) state, an
overall energy $\omega_R$ has been transferred to the $R$ bath and
$\omega_L$ has been transferred to the left bath. Note that unlike the Markov case, we
separately count the energy dissipated at each bath. This
probability satisfies the differential equation
\bea
&&\d{\mathcal P_t(1,\omega_L,\omega_R)}{t} = -\frac{1}{2\pi^2}\int_{0}^{t}ds
\mathcal P_s(1,\omega_L,\omega_R)
\intinf \intinf d\omega_1d\omega_2
C_L(\omega_1)C_R(\omega_2)\Re[e^{i(\omega_0-\omega_1-\omega_2)(t-s)}]
\nonumber\\
&&+\frac{1}{2\pi^2}\int_{0}^{t}ds\intinf\intinf d\omega_1 d\omega_2
\mathcal P_s(0, \omega_1,\omega_2)
C_R(\omega_R-\omega_2)C_L(\omega_L-\omega_1)
\Re[ e^{i(\omega_0+\omega_R-\omega_2+\omega_L-\omega_1)(t-s)}],
\label{eq:probNM}
\nonumber\\
\eea
where we used the fact that $C_{\nu}(\omega)$ is a real function. An
analogous equation can be written for the time evolution of the
probability $\mathcal P_t(0,\omega_L,\omega_R)$. We now introduce
two counting fields $\chi_L$ and $\chi_R$, for each reservoir, and
Fourier transform the above equation with respect to these two
fields. Further, we Laplace transform the resulting equation,
$H(z)=\int_0^{\infty}e^{-zt}h(t)dt$. Utilizing Fourier transform and
Laplace transform convolution relations, Eq. (\ref{eq:probNM})
reduces to a linear equation
\beq
|Z(\chi_L,\chi_R,z)\rangle = \frac{1}{z - \hat W(\chi_L,\chi_R,z)}
\left(z|Z(\chi_L,\chi_R,z\rangle\right)_{z\rightarrow \infty}, 
\label{eq:ZNM} \eeq
with the initial value theorem invoked, $h(t=0)=\lim_{z \to \infty} \ z H(z)$.
The vector $\ket Z$ is defined by
\bea \ket{Z(\chi_L,\chi_R,z)} \equiv
\begin{pmatrix}
\int_0^{\infty}dte^{-zt}\intinf \mathcal
P_t(0,\omega_L,\omega_R)e^{i\omega_L\chi_L}e^{i\omega_R\chi_R}\,d\omega_Ld\omega_R\\
\int_0^{\infty}dte^{-zt}\intinf \mathcal
P_t(1,\omega_L,\omega_R)e^{i\omega_L\chi_L}e^{i\omega_R\chi_R}\,d\omega_Ld\omega_R
\end{pmatrix}
\label{eq:zNM} \eea
and the kernel $\hat W$ represents the matrix
\bea
\hat W(\chi_L,\chi_R,z) =
\begin{pmatrix}
-\gamma^+(z) & \alpha^-(\chi_L,\chi_R,z) \\
\alpha^+(\chi_L,\chi_R,z) & -\gamma^-(z)  \\
\end{pmatrix}
\label{eq:WNM}
\eea
with the elements
\bea \alpha^+(\chi_L,\chi_R,z) &=& \frac{1}{2\pi^2} \intinf \intinf
e^{i\omega_1\chi_L}  e^{i\omega_2\chi_R}
C_R(\omega_2)C_L(\omega_1)\frac{z}{z^2 + (\omega_0 + \omega_1 +
\omega_2)^2}d\omega_1 d\omega_2
\nonumber\\
\alpha^-(\chi_L,\chi_R,z) &=& \frac{1}{2\pi^2}\intinf\intinf
e^{i\omega_1\chi_L} e^{i\omega_2\chi_R}
C_R(\omega_2)C_L(\omega_1)\frac{z}{z^2 + (\omega_0 - \omega_1 -
\omega_2)^2} d\omega_1 d\omega_2
\nonumber\\
\gamma^+(z) &=& \frac{1}{2\pi^2}\intinf\intinf
C_R(\omega_2)C_L(\omega_1) \frac{z}{z^2 + (\omega_0 + \omega_1 +
\omega_2)^2} d\omega_1 d\omega_2
\nonumber\\
\gamma^-(z) &=& \frac{1}{2\pi^2}\intinf\intinf
C_R(\omega_2)C_L(\omega_1) \frac{z}{z^2 + (\omega_0 - \omega_1 -
\omega_2)^2} d\omega_1 d\omega_2
\nonumber\\
\label{eq:MNM}
\eea
We are interested in the total probability, to occupy either states,
\bea
\mathcal
P_t(\omega_L,\omega_R)\equiv\sum_{n=0,1}P_t(n,\omega_L,\omega_R).
\eea
We express it in terms of the characteristic function $e^{S(\chi_L,\chi_R,t)}$,
\bea e^{S(\chi_L,\chi_R,t)}\equiv \intinf \mathcal
P_t(\omega_L,\omega_R)e^{i\omega_L\chi_L}e^{i\omega_R\chi_R}d\omega_Ld\omega_R.
\label{eq:SNM} \eea
This expression generalizes  Eq. (\ref{eq:Gd2}) to the non-markovian
case. It can be formally expressed by an inverse Laplace transform of Eq.
(\ref{eq:ZNM}) \cite{Antti},
\bea e^{S(\chi_L,\chi_R,t)}= \frac{1}{2\pi
i}\int_{a-i\infty}^{a+i\infty}dz e^{zt} \langle I|\frac{1}{z - \hat
W(\chi_L,\chi_R,z)}\left(zZ(\chi_L,\chi_R,z)
\rangle\right)_{z\rightarrow \infty}. \label{eq:GNM} \eea
Here $a$ is a real number, larger than the real part of all the
singularities of the integrand. Equation (\ref{eq:GNM}) is a formal
result. In practice, it is evaluated as follows: First, we note that
the stationary solution (assumed to be unique) of Eq.
(\ref{eq:probNM}) is given by the eigenvector corresponding to the
zero eigenvalue of $\hat W$,
\beq \hat W(\chi_L=0,\chi_R = 0,z = 0)|Z_{SS}\rangle =0. \eeq
Furthermore, as a result of the normalization and conservation of
the total spin probabilities, an eigenvalue of $\hat W$ satisfies
$\lambda_0(\chi_L=0,\chi_R=0,z)=0$, for all $z$ \cite{Antti}.
This can be directly verified in our case, Eq. (\ref{eq:MNM}):
The element  $\alpha^{+}$ ($\alpha^-$) reduces to $\gamma^{+}$ ($\gamma^-$)
when $\chi_{L,R}=0$, and a zero eigenvalue sustains,
irrespective of the value of $z$.
At finite value for the counting fields an eigenvalue
$\lambda_0(\chi_L,\chi_R,z)$ adiabatically develops from this zero
eigenvalue, with small $\chi_L$, $\chi_R$ and $z$. The long time
behavior of the characteristic function is therefore determined by the
pole structure of $(z-\lambda_0(\chi_1,\chi_2,z))^{-1}$ close to zero. This pole
$z_0(\chi_L,\chi_R)$ solves
\bea z_0=\lambda_0(\chi_L,\chi_R,z_0), \label{eq:z0NM} \eea
and it should reduce to $z_0(\chi_L=0,\chi_R=0)=0$, describing the
stationary state. Since all other singularities have  larger negative real
parts, this pole determines the long time behavior of the
characteristic function as
\bea e^{S(\chi_L,\chi_R,t)}\rightarrow
f(\chi_L,\chi_R,z_0)e^{z_0(\chi_L,\chi_R)t}. \label{eq:limitNM} \eea
In the markovian limit $\lambda_0$ does not depend on the $z$
variable, thus trivially
$z_0(\chi_L,\chi_R)=\lambda_0(\chi_L,\chi_R)$.
%

Concluding, the scheme to obtain the CGF proceeds as
follows \cite{Antti}: (i) We  obtain $\lambda_0(\chi_L,\chi_R,z)$,
the eigenvalue of $\hat W(\chi_L,\chi_R,z)$ that adiabatically
develops from the zero (stationary) eigenvalue. (ii) We solve Eq.
(\ref{eq:z0NM}) and obtain the pole $z_0(\chi_L,\chi_R)$. (iii) We
identify the CGF, the analog  of Eq. (\ref{eq:Gd2}), by the pole,
\bea G(\chi_L,\chi_R)\equiv z_0(\chi_L,\chi_R). \eea Back to
(\ref{eq:WNM}), we resolve the eigenvalue
\bea \lambda_0(\chi_L,\chi_R,z) = -\frac{\gamma^+ + \gamma^-}{2} +
\frac{\sqrt{(\gamma^+ - \gamma^-)^2  +4 \alpha^+\alpha^-}}{2},
\label{eq:lamNM} \eea
satisfying
$\lambda_0(\chi_L=0,\chi_R=0,z) = 0$ for all $z$ \cite{Antti}. The
elements of Eq. (\ref{eq:lamNM}) all depend on the variable $z$,
$\alpha^{\pm}$ further depend on the counting fields. In principle,
we should now solve  Eq. (\ref{eq:z0NM}) in order to gain the
CGF, thus the current and its cumulants.

\subsection{Heat Current}
In the long time limit the combination of Eqs. (\ref{eq:GNM}) and
(\ref{eq:limitNM}) leads to
\bea z_0(\chi_L,\chi_R)
\rightarrow \
\frac{1}{t}\ln \intinf
\mathcal
P_t(\omega_L,\omega_R)e^{i\omega_L\chi_L}e^{i\omega_R\chi_R}d\omega_Ld\omega_R.
\eea
It is argued in Ref. \cite{Antti} that
the current and its cumulants can be obtained directly
through the analysis of $\lambda_0$ itself, Taylor expanded around
$z=0$, $\chi_{L}=0$, and $\chi_R=0$,
\bea
\lambda_0(\chi_L,\chi_R,z)=\sum_{n,m,l}\frac{(i\chi_L)^n}{n!}\frac{(i\chi_R)^m}{m!}\frac{z^l}{l!}c^{(n,m,l)},
\eea
with
\bea
c^{(n,m,l)}=\partial^n_{(i\chi_L)}\partial^m_{(i\chi_R)}\partial^l_{z}\lambda_{0}(\chi_L,\chi_R,z)|_{\chi_L,\chi_R,z\rightarrow
0}. \eea
The thermal current, calculated by counting energy flow at the $L$ contact, is given by
\bea &&\langle J_L\rangle \equiv \frac{\avg{\omega_L}_{t}}{t} =
c^{(1,0,0)}. \eea
%
%
Similarly, the current detected at the $R$ contact satisfies
\bea \langle J_R\rangle \equiv \frac{\avg{\omega_R}_{t}}{t} =
c^{(0,1,0)},
\eea
or explicitly
\bea \avg{J_R} =(\gamma^+ + \gamma^-)^{-1}
\left[\alpha^-\frac{\partial \alpha^+}{\partial (i\chi_R)} +
\alpha^+\frac{\partial \alpha^-}{\partial (i\chi_R)}\right]
\Bigg|_{\chi_L,\chi_R,z=0}.\eea
It can be easily verified that this two quantities are identical (with opposite sign),
and equivalent to the markovian result (\ref{eq:JAA}).
It is thus significant to note that our formalism provides a general expression for the energy current,
for many-body systems satisfying the dynamics (\ref{eq:integro}), irrespective of the details of the thermal reservoirs.
One can similarly calculate high order cumulants, by evaluating high
order $c$ terms \cite{Antti}.

\subsection{Fluctuation Theorem}

The assumption of no memory enforces a strict energy conservation
condition for processes between the system and the environments. In
contrast, in the non-markovian regime there is no such an
energy-conservation statement, thus it is not obvious that the
general FT symmetry (\ref{eq:FT}) still holds for any time interval $t$
\cite{Hanggi-rev,Tasaki}. We now prove that the eigenvalue
$\lambda_0(\chi_L,\chi_R,z)$ satisfies the symmetry relation
\bea
\lambda_0(\chi_L,\chi_R,z)=\lambda_0(i\beta_L-\chi_L,i\beta_R-\chi_R,z).
\label{eq:Gn1} \eea
Only in the markovian limit the symmetry is given in terms of the
affinity, as $\lambda_0(\chi)=\lambda_0(i\Delta \beta-\chi)$. Since
the counting fields and $z$ are independent variables, these
symmetry relations translate into the analogous relations for the
CGF itself, $z_0(\chi_L,\chi_R)$. This result exemplifies that while
microreversibility is sufficient for deriving the basic symmetry
relation (\ref{eq:Gn1}), the SSFT holds only under more restrictive
conditions, dictated here by the bath relaxation timescale
\cite{Mukamel-rev,Tasaki}.

The symmetry of $\lambda_0(\chi_L,\chi_R,z)$, thus the symmetry of
the CGF, is coded in the product of terms that depend on
the counting fields, $\mathcal D(\chi_L,\chi_R,z)\equiv
\alpha^+\alpha^-$, see Eq. (\ref{eq:lamNM}). We can readily confirm that
\bea && \alpha^+(i\beta_L-\chi_L,i\beta_R-\chi_R,z) = \nonumber\\
&& \frac{1}{2\pi^2}\intinf\intinf
e^{-i\omega_1\chi_L}e^{-\beta_L\omega_1}e^{-i\omega_2\chi_R}e^{-\beta_R\omega_2}
C_R(\omega_2)C_L(\omega_1)\frac{z}{z^2 + (\omega_0 + \omega_1 +
\omega_2)^2}d\omega_1 d\omega_2 \nonumber\\
&&=\alpha^-(\chi_L,\chi_R,z). \eea
This result is obtained by changing variables,
$\omega_1\rightarrow -\omega_1$ and $\omega_2\rightarrow -\omega_2$, then
utilizing the detailed balance relation,
$C_{\nu}(-\omega)=C_{\nu}(\omega)e^{-\beta_{\nu}\omega}$. Similarly,
it can be proved that
\bea && \alpha^-(i\beta_L-\chi_L,i\beta_R-\chi_R,z) =
\alpha^+(\chi_L,\chi_R,z). \eea
As a result, the symmetry relation (\ref{eq:Gn1}) is confirmed, and
the CGF, reached by solving Eq.
(\ref{eq:lamNM}),  similarly satisfies
\bea z_0(\chi_L,\chi_R)=z_0(i\beta_L-\chi_L,i\beta_R-\chi_R).
\eea
The probability itself is given by an inverse Fourier transform [Eq.
(\ref{eq:SNM})] with respect to both fields,
\bea {\mathcal
P}_{t}(\omega_L,\omega_R)=\frac{1}{(2\pi)^2}\int_{-\infty}^{\infty}
\intinf e^{
z_0(\chi_L,\chi_R)t}e^{-i\omega_L\chi_L}e^{-i\omega_R\chi_R}
d\chi_Ld\chi_R. \eea
Based on the symmetry of the CGF, it can be readily proved that in the long time limit
the following ``basic" fluctuation relation holds \cite{Hanggi-rev}
\beq \frac{\mathcal P_t(\omega_L,\omega_R)}{\mathcal
P_t(-\omega_L,-\omega_R)} = e^{\beta_L\omega_L}e^{\beta_R\omega_R}.
\eeq
The ``standard" fluctuation theorem, expressed in terms of the affinity $\Delta \beta=\beta_R-\beta_L$
is regained
when the kernel $\hat W$ reduces to the markovian result; The two
counting fields then trivially count the same amount of energy,
$\omega_L=-\omega_R$. This can be explicitly shown by evaluating the elements
$\alpha^+$ and $\alpha^-$ in the markovian limit $z=0$. We find that
\bea \alpha^+(\chi_L,\chi_R,z=0) &=& \frac{1}{2\pi} \intinf
e^{-i(\omega_0+\omega_2)\chi_L}  e^{i\omega_2\chi_R}
C_R(\omega_2)C_L(-\omega_0-\omega_2)d\omega_2
\nonumber\\
\alpha^-(\chi_L,\chi_R,z=0) &=& \frac{1}{2\pi}\intinf
e^{i(\omega_0-\omega_2)\chi_L} e^{i\omega_2\chi_R}
C_R(\omega_2)C_L(\omega_0-\omega_2)d\omega_2
\nonumber\\
\eea
We now define a new counting field, $\chi=\chi_R-\chi_L$, and
immediately verify that the product of these two objects, $\mathcal D$, satisfies
\bea \mathcal D(\chi)=\mathcal D(i\Delta\beta-\chi). \eea
This directly implies on the same symmetry for the markovian CGF,
\beq \lambda_0(\chi_R-\chi_L) = \lambda_0(i\Delta \beta-
\chi_R+\chi_L), \eeq
leading to the standard heat exchange SSFT, Eq. (\ref{eq:SSFTM}).

One should note that Eq. (\ref{eq:limitNM}) has already involved the
assumption of long times, such that only one eigenvalue of $\hat W$,
with the smallest (absolute) real value, dictates the dynamics. The
$z$ dependence in Eq. (\ref{eq:MNM}) thus manifests itself when the
bath decorrelation time is long, comparable with the inverse
relaxation rates of the subsystem. This observation establishes the
regime of validity of the SSFT, Eq. (\ref{eq:FT}). It holds when the
interval $t$ is long, beyond the bath memory time. We recall that
for strictly harmonic systems one directly obtains the SSFT
\cite{Dhar}, without any reference to the bath characteristic
timescale. This is because in coherent systems the reservoirs only
serve as a source for excitations, which then elastically cross the
impurity. In contrast, in the present model inelastic bath-induced
processes are involved in the energy transfer process, making the
bath decorrelation time a relevant parameter for the dynamics.


\section{Conclusions}

We presented here a scheme for obtaining the CGF, thus the current
and its moments for the non-equilibrium spin-boson model, an eminent
many-body impurity model. A heat exchange SSFT was established for
quantum systems incorporating strong system-bath interactions and
anharmonic effects. Our derivation relays on the NIBA equations,
originally developed for the equilibrium spin-boson model,
generalized to describe the dynamics of a spin impurity coupled to
multiple thermal reservoirs. Our study provides closed expressions
for the CGF, useful for deriving the distribution of heat
fluctuations, the averaged current and the thermal noise power. We
also showed explicitly that the timescale controlling the onset of
the SSFT is the decorrelation time of the reservoirs. Future work
will be devoted to generalizing our study to systems showing
coherence effects, either using path integral formulation, or
quantum master equation methods. Exploring the analogous dynamics
for a fermionic system under voltage and temperature biases will be
the topic of future studies.

\begin{acknowledgments}
The research of YN was funded by the Early Research Award of DS. DS
further acknowledges support from NSERC discovery grant.
\end{acknowledgments}


\renewcommand{\theequation}{A\arabic{equation}}
\setcounter{equation}{0}  
\section*{Appendix A: Derivation of EOM for the resolved probability}

The equation of motion for the resolved probabilities in the markovian limit, Eq. (\ref{eq:Pw}), are explained here,
based on the population dynamics (\ref{eq:EOMsm}). For clarity, we include
this equation again,
\bea
\dot p_1(t)=-p_1(t)C(\omega_0) + p_0(t) C(-\omega_0).
\label{eq:Apop}
\eea
The population of each state at time $t$ can be expressed in terms
of the resolved probability $\mathcal P_t(n,\omega)$, $n=0,1$, that
within the time $t$ a total of energy $\omega$ has been transferred
from the left bath to the right bath, while the spin is populating
the $n$ ($n=0,1$) state at time $t$,
\bea
p_1(t) = \intinf\mathcal P_t(1,\omega)d\omega,
\nonumber \\
p_0(t) = \intinf \mathcal P_t(0,\omega)d\omega.
\eea
Plugging these integrals in the dynamical equation (\ref{eq:Apop}),
it becomes (ignoring $1/2\pi$ factors for simplicity)
\bea
\d{}{t}\intinf \mathcal P_t(1,\omega)d\omega & = & -\int_{-\infty}^\infty C_L(\omega_0 - \omega_1)C_R(\omega_1)d\omega_1 \times
\intinf \mathcal P_t(1,\omega_2)d\omega_2
\nonumber \\
& + & \int_{-\infty}^\infty C_L(-\omega_0 -
\omega_1)C_R(\omega_1)d\omega_1 \times \intinf \mathcal
P_t(0,\omega_2)d\omega_2. \label{eq:Ares1} \eea
We now equate identical energy terms, thus get the resolved dynamics (\ref{eq:Pw})
\bea \frac{d\mathcal P_t(1,\omega)}{dt}&=& -\mathcal P_t(1,\omega)
\intinf C_L(\omega_0-\omega_1)C_R(\omega_1)d\omega_1
\nonumber\\
& +&\intinf \mathcal P_t(0,\omega_1)
C_L(-\omega_0-\omega+\omega_1)C_R(\omega-\omega_1) d\omega_1.
\label{eq:Ares2}
\eea
For further validating this equation, we attempt to recover Eq.
(\ref{eq:Apop}) by integrating this equation over frequency. The
first term in Eq. (\ref{eq:Ares2}) trivially reduces to the first
term in (\ref{eq:Ares1}). The second term in Eq. (\ref{eq:Ares1}) is
restored following a variable change,
\bea &&\intinf d\omega \intinf \mathcal P_t(0,\omega_1)
C_L(-\omega_0-\omega + \omega_1) C_R(\omega-\omega_1) d\omega_1=
\nonumber\\
&& \intinf d\omega_2 C_L(-\omega_0-\omega_2)C_R(\omega_2) \intinf d\omega  \mathcal P_t(0,\omega-\omega_2) =
 C(-\omega_0)p_0(t)
\eea
%

\renewcommand{\theequation}{B\arabic{equation}}
\setcounter{equation}{0}  
\section*{Appendix B: Derivation of the current and its noise power in the markovian limit}

We derive here the steady-state heat current, Eq. (\ref{eq:J2}), and
the noise power Eq. (\ref{eq:S2}). We begin by solving the kinetic
equations (\ref{eq:EOMsm}) in the long-time limit. The steady-state
populations are given by
\bea p_1 = \frac{k_u}{k_u + k_d} = \frac{C(-\omega_0)} {C(-\omega_0)
+ C(\omega_0)}
\nonumber \\
p_0 = \frac{k_d}{k_u + k_d}= \frac{C(\omega_0)}{C(-\omega_0) +
C(\omega_0)}.
\label{eq:popAA}
\eea
We now study the first derivative of the CGF, Eq. (\ref{eq:G}),
with respect to the counting field,
\bea \langle J\rangle  = \frac{dG(\chi)}{d(i\chi)}\Bigg|_{\chi=0} &
= & \left[\d{C^d(\chi)}{i\chi}C^u(\chi)+
\d{C^u(\chi)}{i\chi}C^d(\chi)\right]
\nonumber\\
&\times& \left[ (C(\omega_0) - C(\omega_0))^2 +
4C^d(\chi)C^u(\chi)\right]^{-1/2}\Big|_{\chi=0}.
\label{eq:JB} \eea
Note that $C^d(\chi=0)=C(\omega_0)$ and $C^u(\chi=0)=C(-\omega_0)$,
a direct result of Eqs. (\ref{eq:Cdu}) and (\ref{eq:Cdu2}). We
identify the second term in the expression above by the sum
$(C(\omega_0)+C(-\omega_0))^{-1}$. The partial derivatives  are
given by (ignoring $(2\pi)^{^-1}$ factors for simplicity)
\bea
\frac{dC^{u}(\chi)}{d{(i\chi)}}\Big|_{\chi=0}&=&\intinf \omega C_R(\omega)C_L(-\omega-\omega_0) d\omega= -\intinf \omega C_R(-\omega)C_L(\omega-\omega_0) d\omega
\nonumber\\
\frac{dC^{d}(\chi)}{d{(i\chi)}}\Big|_{\chi=0}&=& \intinf \omega C_R(\omega)C_L(\omega_0-\omega)d\omega.
\eea
Plugging these terms into Eq. (\ref{eq:JB}) we find the current
\bea \langle J\rangle &=& \frac{1}{C(\omega_0)+C(-\omega_0)} \Big[
 C(-\omega_0) \intinf \omega C_R(\omega)C_L(\omega_0-\omega)d\omega
\nonumber \\
 & - &  C(\omega_0) \intinf \omega C_R(-\omega)C_L(\omega-\omega_0)d\omega
 \Big].
 \label{eq:JAA}
\eea
It is significant to note that this expression stays intact for
non-markovian systems \cite{Antti}. Next we adopt the steady-state
population (\ref{eq:popAA}) and simplify the result,
\bea \avg J=    p_1\intinf\omega
C_R(\omega)C_L(\omega_0-\omega)d\omega -p_0\intinf  \omega
C_R(-\omega)C_L(\omega-\omega_0) d\omega, \eea
which is precisely Eq. (\ref{eq:J2}).
We now verify that $ \langle J(\Delta T)\rangle =-\langle J(-\Delta T)\rangle$ for a spatially symmetric system.
Upon exchange of the temperature polarity, the above expression becomes
\bea \langle J(-\Delta T)\rangle=    p_1\intinf\omega
C_L(\omega)C_R(\omega_0-\omega)  d\omega - p_0\intinf  \omega
C_L(-\omega)C_R(\omega-\omega_0)d\omega. \eea
We change variables, $\omega-\omega_0=-x$,  and get
\bea \langle J(-\Delta T)\rangle=    p_1\intinf(\omega_0-x)
C_L(\omega_0-x)C_R(x)  dx- p_0\intinf  (\omega_0-x)
C_L(x-\omega_0)C_R(-x)dx \nonumber\\ \eea
This expression can be organized as
\bea
 \langle J(-\Delta T)\rangle&=&  \omega_0
 \left[p_1\intinf   C_L(\omega_0-x)C_R(x)dx  -
p_0\intinf C_L(x-\omega_0)C_R(-x)  dx \right]
\nonumber\\
&-&  \left[p_1\intinf  x C_L(\omega_0-x)C_R(x)dx  - p_0\intinf x
C_L(x-\omega_0)C_R(-x)  dx \right]. \eea
Since the first line fade away once combining the definition
(\ref{eq:Cw0}) and the steady-state population (\ref{eq:SSpop}), we
establish the odd symmetry for the current with $\Delta T$. The
 noise power is formally given by
\bea && \langle S\rangle  =
\frac{d^2G(\chi)}{d(i\chi)^2}\Bigg|_{\chi=0} \nonumber\\
&& =  \left[ \frac{d^2C^d(\chi)}{d(i\chi)^2} C^u(\chi)+
\frac{d^2C^u(\chi)}{d(i\chi)^2} C^d(\chi)+
2\d{C^u(\chi)}{i\chi}\d{C^d(\chi)}{i\chi}\right]
\nonumber\\
&&\times \left[ (C(\omega_0) - C(-\omega_0))^2 +
4C^d(\chi)C^u(\chi)\right]^{-1/2}\Big|_{\chi=0}
\nonumber\\
&&- 2\left[ \frac{dC^d(\chi)}{d(i\chi)} C^u(\chi)+
\frac{dC^u(\chi)}{d(i\chi)} C^d(\chi) \right]^2 \left[ (C(\omega_0)
- C(-\omega_0))^2 + 4C^d(\chi)C^u(\chi)\right]^{-3/2}\Bigg|_{\chi=0}
\label{eq:S}. \eea
Using the explicit expressions for the correlations we reduce it to

\bea \langle S \rangle  = &
&p_1\int_{-\infty}^{\infty}\omega^2C_R(\omega)C_L(\omega_0-\omega)d\omega
+
p_0\int_{-\infty}^{\infty}\omega^2C_R(-\omega)C_L(\omega-\omega_0)d\omega+
\nonumber\\
&-& 2\frac{1}{C(\omega_0)+C(-\omega_0)}\int_{-\infty}^{\infty}\omega
C_R(-\omega)C_L(\omega-\omega_0)d\omega\int_{-\infty}^{\infty}\omega
C_R(\omega)C_L(\omega_0-\omega)d\omega+
\nonumber\\
&-& 2\frac{1}{C(\omega_0)+C(-\omega_0)}\avg{J}^2. \eea
%

\renewcommand{\theequation}{C\arabic{equation}}
\setcounter{equation}{0}  
\section*{Appendix C: The spin-boson model in the weak coupling limit}


We study here the counting statistics of the {\it unbiased}
($\omega_0=0$) spin-boson model, and verify the validity of the SSFT
in this case, both under the Born-Markov Approximation
\cite{Mukamel-weak,HanggiBerry}. Our starting point is the SB
Hamiltonian [Eq. (\ref{eq:HSB})]. We take $\omega_0=0$ and apply a
unitary transformation
\bea
U^{\dagger}\sigma_z U=\sigma_x, \,\,\,\,\,  U^{\dagger}\sigma_x U=\sigma_z
\eea
with $U=\frac{1}{\sqrt{2}}(\sigma_x+\sigma_z)$, to obtain the
transformed Hamiltonian $H_W=U^{\dagger}HU$,
\bea
H_W &=& H_0 + H_I + H_B \nonumber\\
H_0&=&\frac{\Delta}{2} \sigma_z ; \,\,\, H_I= \sigma_x
\sum_{\nu,j}\lambda_{j,\nu}(b_{j,\nu}^\dagger + b_{j,\nu})
\nonumber\\ H_B&=&\sum_{\nu}H_{\nu}; \,\,\, H_{\nu}=
\sum_{j}\omega_j b_{j,\nu}^{\dagger}b_{j,\nu}. \label{eq:HSBw} \eea
Note that the subsystem energy gap is now given by $\Delta$, rather
than $\omega_0$ as in the original spin-boson description  [Eq.
(\ref{eq:HSB})]. This form is a convenient starting point for a
perturbation-theory calculation, assuming weak system-bath coupling.
We outline next the principles of this standard approach
\cite{Weiss}. Details, for the two-bath scenario, can be found in
Ref. \cite{SegalM}. We begin with an equation of motion for the
total density matrix $\rho(t)$ in the interaction representation,
\bea \dot \rho = -i[H_I(t),\rho(t)]. \eea
The operators  are given by
$O(t)=e^{i(H_0+H_B)t}O e^{-i(H_0+H_B)t}$. We now make the following
assumptions: (i) At the initial time the reservoirs are (separately)
maintained in thermal equilibrium, isolated from the subsystem, and
(ii) the spin and the baths are weakly coupled, allowing for a
weak-coupling expansion with respect to the interaction term in  Eq.
(\ref{eq:HSBw}). This results in
\bea \dot \rho_S(t)=-\int_0^{t} ds {\rm Tr} [
H_I(t),[H_I(s),\rho_S(s) \otimes \rho_B ] ], \eea
which is a {\it non-markovian} equation of motion. Here
$\rho_S(t)={\rm Tr}[\rho(t)]$ is the reduced density matrix
of the system; the trace is performed over the two reservoirs. The
bath density matrix is a product state, $\rho_B=\rho_L \otimes
\rho_R$, of the two canonical density matrices
 $\rho_{\nu}=e^{-H_{\nu}/T_{\nu}}/{\rm Tr}_{\nu}[e^{-H_{\nu}/T_{\nu}}]$.
In our model (\ref{eq:HSBw}) the population dynamics becomes
decoupled from the coherences dynamics. It obeys
\bea \dot p_1 = -2\Re \int_0^t e^{i\Delta (t-s)} g(t-s) p_1(s)ds +
2\Re \int_0^t e^{-i\Delta(t-s)}g(t-s)p_0(s)ds.
\label{eq:popW}
\eea
Here $p_{n}(t)=(\rho_S(t))_{n,n}$ ($n=0,1$) and
$g(\tau)=\sum_{\nu}g_{\nu}(\tau)$, with
\bea g_{\nu}(\tau)=\int_0^{\infty} \frac{J_{\nu}(\omega)}{4\pi}
\left[n_{\nu}(\omega)e^{i\omega\tau} +
(n_{\nu}(\omega)+1)e^{-i\omega\tau}\right]d\omega. \eea
The bath spectral function is given by
$J_{\nu}(\omega)=4\pi\sum_{j}\lambda_{j,\nu}^2\delta(\omega-\omega_j)$
and the function $n_{\nu}(\omega) =[e^{\beta_{\nu}\omega}-1]^{-1}$
denotes the Bose-Einstein distribution. We now make the markovian
approximation, assuming that bath correlations decay on a time scale
shorter than the subsystem characteristic timescale. This converts
Eq. (\ref{eq:popW}) into a kinetic-Master equation,
\bea
&&\dot p_1 = - p_1\sum_{\nu}k_{1\rightarrow 0}^{\nu} + p_0\sum_{\nu}k_{0\rightarrow 1}^{\nu}  \nonumber\\
&&p_1(t)+p_0(t)=1,
\label{eq:pd}
\eea
where  the Fermi-golden rule transition rates are evaluated at the
subsystem energy gap $\Delta$, satisfying
\bea k_{0\rightarrow 1}^{\nu}= \Gamma_{\nu}(\Delta)n_{\nu}(\Delta),
\,\,\,\,\,\,\, k_{1\rightarrow 0}^{\nu}=
\Gamma_{\nu}(\Delta)[1+n_{\nu}(\Delta)]. \eea
The rate $\Gamma_{\nu}(\omega)=2\pi \sum_{j}\lambda_{j,\nu}^2
\delta(\omega_j-\omega)$ denotes the temperature independent part of
the relaxation rate. The dynamics (\ref{eq:pd}) describes spin flip
processes accompanied by an energy transfer at the amount of
$\Delta$ to either the left or the right reservoirs.

We proceed and derive the cumulant generating function in the
present weak coupling limit following \cite{HanggiBerry}. We begin
by defining $\mathcal P_t(n,q\Delta)$ as the probability that within
the time $t$ a total energy $q \Delta$ has been transferred from the
left bath to the right bath, while the spin is populating the $n$
($n=0,1$) state at time $t$. Note that $q$ here is an integer, since
energy is transferred here in discrete quanta of $\Delta$, between
the two baths. In other words, whenever the spin flips, the spin gap
$\Delta$ is dissipated or absorbed at either the left or the right
reservoir. Thus,
\bea \frac{d \mathcal P_t(0,q\Delta)}{dt}& = & - \mathcal
P_t(0,q\Delta) (k_{0\rightarrow 1}^L +  k_{0\rightarrow 1}^R ) +
\mathcal P_t(1,(q-1)\Delta) k_{1\rightarrow 0}^R +  \mathcal
P_t(1,q\Delta) k_{1\rightarrow 0}^L
\nonumber\\
\frac{d \mathcal P_t(1,q\Delta)}{dt}& = & - \mathcal P_t(1,q\Delta)
(k_{1\rightarrow 0}^L +  k_{1\rightarrow 0}^R ) + \mathcal
P_t(0,(q+1)\Delta) k_{0\rightarrow 1}^R +  \mathcal P_t(0,q\Delta)
k_{0\rightarrow 1}^L.
\label{eq:Pwweak}
\eea
We Fourier transform these equations with the counting field $\chi$
to obtain the characteristic function,
\bea \ket{Z(\chi,t)} \equiv
\begin{pmatrix}
\sum_q \mathcal P_t(0,q\Delta)e^{iq\Delta\chi}\
\\ \sum_q \mathcal
P_t(1,q\Delta)e^{iq\Delta\chi}\
\end{pmatrix}
\label{eq:zW} \eea
satisfying a first order differential equation,
\beq \d{\ket{Z(\chi, t)}}{t} = - \hat W(\chi)\ket{Z(\chi, t)},
\label{eq:ZW} \eeq
with the matrix
\bea \hat W=
\begin{pmatrix}
k_{0\rightarrow 1}^L+k_{0\rightarrow 1}^R  & -k_{1\rightarrow 0}^L - k_{1\rightarrow 0}^R e^{i\chi\Delta}   \\
-k_{0\rightarrow 1}^L -k_{0\rightarrow 1}^R e^{-i\chi \Delta}  &   k_{1\rightarrow 0}^L + k_{1\rightarrow 0}^R \\
\end{pmatrix}
\label{eq:muweak}
\eea
The CGF is given by the negative of the smallest eigenvalues of this
matrix,
\bea G(\chi)=\frac{-A + \sqrt{A^2+4B(\chi)}}{2}. \eea
The coefficients are defined as
\bea A&=&\Gamma_L[1+2n_L(\Delta)]+
\Gamma_R[1+2n_R(\Delta)],
\nonumber\\
B(\chi)&=& \Gamma_L\Gamma_Rn_L(\Delta)n_R(\Delta) \left[
(e^{-i\chi\Delta}-1) e^{\beta_L\Delta} +  (e^{i\chi \Delta}-1)
e^{\beta_R\Delta}\right]. \eea
For brevity, we have discarded the direct dependence of the rates on
frequency, $\Gamma(\Delta)$. It can be easily verified that the
cumulant generating function satisfies the symmetry
$G(\chi)=G(i\Delta \beta- \chi)$ with $\Delta\beta=\beta_R-\beta_L$.
This symmetry can be translated into the fluctuation relation at
long time $t$ \cite{Mukamel-rev,HanggiBerry},
\bea \frac{{\mathcal P}_{t}(q\Delta)}{{\mathcal
P}_{t}(-q\Delta)}=e^{q \Delta (\beta_R-\beta_L)}. \eea
Comparing this result to the strong coupling expression
(\ref{eq:SSFTM}), we note that in the unbiased-weak coupling limit
the discrete energy $q\Delta$ replaces the continuous variable
$\omega$, since the reservoirs here accept or contribute energy in
quanta of the spin spacing $\Delta$. Finally, the current and the
noise power are given by
\bea
\langle J \rangle&=& \frac{1}{A} \frac{\partial B}{\partial i\chi}\Big|_{\chi=0}
\nonumber\\
\langle S \rangle &=& \frac{1}{A} \left[ \frac{\partial^2 B
}{\partial (i\chi)^2} - \frac{2}{A^2} \left( \frac{\partial
B}{\partial i\chi}\right)^2\right]\Bigg|_{\chi=0}. \eea
The elements in this expression are
\bea &&\frac{\partial B}{\partial (i\chi)}\Big|_{\chi=0}=
\Delta\Gamma_L\Gamma_R  [n_L(\Delta)-n_R(\Delta)], \,\,\,
\nonumber\\
&& \frac{\partial^2 B}{\partial (i\chi)^2} \Big|_{\chi=0}=
-\Delta^2\Gamma_L\Gamma_R
  [n_L(-\Delta)n_R(\Delta) + n_R(-\Delta)n_L(\Delta)].
\eea
We find that in this weak coupling limit the current satisfies
\bea \avg J =\Delta\frac{\Gamma_L\Gamma_R
[n_L(\Delta)-n_R(\Delta)]}{\Gamma_L[1+2n_L(\Delta)]+
\Gamma_R[1+2n_R(\Delta) ]}.
\label{eq:ACJ}
 \eea
This result agrees with previous studies \cite{Rectif}.

\end{document}